\documentclass[
    aps,
    prx,
    10pt,
    longbibliography,
    amsfonts,
    amssymb,
    amsmath,
    superscriptaddress,
    twocolumn,
    floatfix 
]{revtex4-2}

\usepackage{xcolor}

\definecolor{qubo_red}{HTML}{C73626}
\definecolor{qubo_blue}{HTML}{215CAF}
\definecolor{qubo_green}{HTML}{3A994E}
\definecolor{qubo_orange}{HTML}{FFAE4F}

\usepackage{graphicx}
\usepackage{physics}
\usepackage[colorlinks]{hyperref}
\hypersetup{
    colorlinks = true,
    linkcolor = qubo_red,
    citecolor = qubo_red,
    filecolor = qubo_blue,      
    urlcolor = qubo_blue,
    pdftitle={Entanglement-assisted heuristic for variational solutions of discrete optimization problems},
}

\usepackage[labelfont=bf, labelsep=period]{caption}

\usepackage{dsfont}
\usepackage{xfrac}
\usepackage{subcaption}
\usepackage{placeins}
\usepackage{amsmath}
\usepackage[normalem]{ulem}
\usepackage[capitalize]{cleveref}

\usepackage{ragged2e}
\DeclareCaptionJustification{justified}{\justifying}
\DeclareMathOperator*{\argmin}{\text{argmin}}
\newcommand{\U}[1]{\mathcal{U}(#1)}
\newcommand{\V}[1]{\mathcal{V}(#1)}

\newcommand{\np}{\texttt{NP}}
\newcommand{\nphard}{\texttt{NP-Hard}}

\newcommand{\hamiltonian}{\hat{\mathcal{H}}}
\newcommand{\pauli}[2]{\hat{\sigma}_{\ensuremath{\mathrm{#1}}}^{(\ensuremath{#2})}}
\newcommand{\sigmax}[1]{\pauli{x}{#1}}
\newcommand{\sigmaz}[1]{\pauli{z}{#1}}

\begin{document}


\title{Entanglement-assisted variational algorithm for discrete optimization problems}

\author{
    \firstname{Lorenzo}
    \surname{Fioroni}
}
\email{lorenzo.fioroni@epfl.ch}

\author{
    \firstname{Vincenzo}
    \surname{Savona}
}
\email{vincenzo.savona@epfl.ch}

\affiliation{Laboratory of Theoretical Physics of Nanosystems, École Polytechnique Fédérale de Lausanne (EPFL), CH-1015 Lausanne, Switzerland}
\affiliation{Center for Quantum Science and Engineering,\\ École Polytechnique Fédérale de Lausanne (EPFL), CH-1015 Lausanne, Switzerland}


\begin{abstract}
    From fundamental sciences to economics and industry, discrete optimization problems are ubiquitous.
    Yet, their complexity often renders exact solutions intractable, necessitating the use of approximate methods.
    Heuristics inspired by classical physics have long played a central role in this domain.
    More recently, quantum annealing has emerged as a promising alternative, with hardware implementations realized on both analog and digital quantum devices.
    Here, we develop a heuristic inspired by quantum annealing, using Generalized Coherent States as a parameterized variational Ansatz to represent the quantum state.
    This framework allows for the analytical computation of energy and gradients with low-degree polynomial complexity, enabling the study of large problems with thousands of spins.
    Concurrently, these states capture non-trivial entanglement, crucial for the effectiveness of quantum annealing.
    We benchmark the heuristic on the three-dimensional Edwards-Anderson model and 
    compare the solution quality and runtime of our method to other popular heuristics.
    Our findings suggest that it offers a scalable way to leverage quantum effects for complex optimization problems, with the potential to complement or improve upon conventional alternatives in large-scale applications.
\end{abstract}
    \maketitle 



\section{Introduction}

Optimization problems play a pivotal role in a wide range of fields.
From optimizing the allocation of medical resources in healthcare systems~\cite{dentonOptimizationSurgerySequencing2007, cardoenOperatingRoomPlanning2010} to improving energy distribution in smart grids~\cite{jordehiOptimisationDemandResponse2019, logenthiranDemandSideManagement2012} and enhancing traffic flow in urban areas~\cite{farahaniReviewUrbanTransportation2013, guihaireTransitNetworkDesign2008}, these tasks aim to identify which input from a pre-defined set minimizes a given metric, referred to as the \emph{loss function}.
A paradigmatic class of optimization problems is that of \emph{Quadratic Unconstrained Binary Optimization} (QUBO)~\cite{kochenbergerUnconstrainedBinaryQuadratic2014}, where the loss function is a quadratic form of binary variables.
Despite their simple formulation, QUBO problems are known to be computationally challenging, as they belong to the \nphard{} complexity class~\cite{barahonaComputationalComplexityIsing1982, vavasisComplexityTheoryQuadratic2001}. 
This means that no algorithm is known to solve an arbitrary QUBO instance in polynomial time, and the required computational resources grow exponentially with the problem size.
At the same time, QUBO problems are of great practical relevance, as any other \np{} problem can be efficiently reduced to them~\cite{lucasIsingFormulationsMany2014}.
Consequently, the development of fast algorithms for approximating the optimal solution of these problems is of great interest.
These can be broadly divided into two categories: \emph{approximation algorithms} and \emph{heuristics}.
Approximation algorithms aim to find a solution that is guaranteed to be up to a certain factor off the optimal one~\cite{christofidesWorstCaseAnalysisNew2022, goemansImprovedApproximationAlgorithms1995}, while heuristics aim to find a good solution in a reasonable amount of time, without any guarantee of optimality~\cite{bianchiSurveyMetaheuristicsStochastic2009}.

Multiple heuristics have been developed to tackle QUBO problems efficiently, often inspired by various areas of physics~\cite{wangCoherentIsingMachine2013, haribaraPerformanceEvaluationCoherent2017, calvanesestrinatiMultidimensionalHyperspinMachine2022, tsukamotoAcceleratorArchitectureCombinatorial2017, quApplicationSpiralEnhanced2024, zeng_performance_2024}. 
For example, simulated annealing~\cite{kirkpatrickOptimizationSimulatedAnnealing1983} emulates the behavior of a system at finite temperature, exploring the configuration space via thermal fluctuations, while simulated bifurcation~\cite{gotoCombinatorialOptimizationSimulating2019} draws inspiration from nonlinear dynamics to navigate the solution landscape efficiently. 
These physics-inspired heuristics have further motivated the development of specialized hardware accelerators, such as the Coherent Ising Machine~\cite{mohseni_ising_2022} and the Fujitsu Digital Annealer~\cite{matsubara2018complex, tsukamoto2017fujitsu, matsubara2020asia}, among others.

Quantum physics offers a promising approach to solving QUBO problems through \emph{quantum annealing}~\cite{finnilaQuantumAnnealingNew1994, santoroOptimizationUsingQuantum2006, dasColloquiumQuantumAnnealing2008, albashDemonstrationScalingAdvantage2018, crossonProspectsQuantumEnhancement2021}.
In quantum annealing, the classical problem is first mapped onto the ground state of a quantum Hamiltonian, which is then obtained via adiabatic state preparation.
Despite being the subject of extensive research, a definitive demonstration of the advantage of quantum annealing over classical heuristics remains an open question.
Numerous studies have been conducted, aimed at simulating its underlying physical process to understand the effects of finite temperature, noise, diabatic transitions, and other deviations from the ideal scenario~\cite{gardasDefectsQuantumComputers2018, mishraFiniteTemperatureQuantum2018, childsRobustnessAdiabaticQuantum2001}. 
The goal of these studies is to describe the quantum state and its dynamics as accurately as possible, to ultimately characterize quantum annealing and identify its differences from classical optimization methods.
To this purpose, a variety of techniques have been adopted, mostly relying either on Path Integral Monte Carlo (PIMC)~\cite{ronnowDefiningDetectingQuantum2014,santoroTheoryQuantumAnnealing2002,martonakQuantumAnnealingPathintegral2002}, or on Variational Monte Carlo (VMC) dynamics.
In the latter case, several advanced variational Ansätze have been used to capture the many-body quantum correlations arising along the evolution~\cite{hibat-allahVariationalNeuralAnnealing2021a,carleoSimulatingAdiabaticQuantum2024}.
While these studies are crucial for understanding the mechanisms of quantum annealing, they are not suitable for use as efficient optimization heuristics due to their high computational demands and unfavorable scaling with problem size. 
Additionally, many of these methods rely on Monte Carlo sampling, which further increases computational overhead.
In contrast, a classical optimization algorithm inspired by quantum annealing might forgo some physical accuracy in favor of greater efficiency.
Recent studies in this direction proposed using a product-state Ansatz to describe the state of the system~\cite{bowlesQuadraticUnconstrainedBinary2022, veszeliMeanFieldApproximation2021}.
This method, referred to as Local Quantum Annealing (LQA), allows simulating the state's evolution analytically, avoiding the need for Monte Carlo sampling.
However, the simplicity of the Ansatz severely limits the range of states it can represent, resulting in a fast but less accurate heuristic.
In particular, it cannot capture any of the entangled states that arise during quantum annealing.

In this work, we propose a quantum-inspired heuristic for solving QUBO problems. 
Drawing inspiration from the quantum annealing process, we develop an efficient variational procedure that emulates its dynamics in a fully analytical way. 
The variational Ansatz we employ is based on Generalized Group-Theoretic Coherent States (GCS)~\cite{guaita_generalization_2021, schindler_variational_2022}, which allow for the efficient evaluation of the energy and its gradient.
At the same time, it captures to some extent the entanglement structure that emerges during the quantum annealing process, thus leveraging the advantage it provides.
Without the need for Monte Carlo sampling, our algorithm is highly scalable and allows for the efficient optimization of problems with thousands of variables.

We benchmark our algorithm on random instances of the three-dimensional Edwards-Anderson model~\cite{barahonaComputationalComplexityIsing1982}, comparing its performance to that of standard heuristics such as Simulated Annealing (SA)~\cite{kirkpatrickOptimizationSimulatedAnnealing1983}, Local Quantum Annealing (LQA)~\cite{bowlesQuadraticUnconstrainedBinary2022}, and Parallel Tempering with Iso-energetic Cluster Moves (PT-ICM)~\cite{bauza_scaling_2024}. 
We identify parameter regimes where our algorithm holds an advantage over the other methods tested, proving that it could serve as a viable alternative to other state-of-the-art classical heuristics.


\section{Results}
    
A QUBO problem is fully specified by a real-valued and symmetric matrix $J$ and a real-valued bias vector $\vb{b}$ through the relation
\begin{equation}
    \vb{z}^\star  = \argmin_{\vb{z} \in \qty{0,\;1}^{N}} \left[\sum_{i,j = 1}^N J_{ij} z_i z_j + \sum_{i=1}^N b_i z_i\right] \text{,}
    \label{eq: binary qubo}
\end{equation}
which also defines its solution $\vb{z}^\star$.
Up to a linear transformation of its variables, the binary optimization problem in Eq.~\eqref{eq: binary qubo} can be mapped onto 
\begin{equation}
    \vb{s}^\star = \argmin_{\vb{s} \in \qty{\pm 1}^{N}} \left[\sum_{i,j = 1}^N W_{ij} s_i s_j + \sum_{i=1}^N c_i s_i\right] \text{,}
    \label{eq: qubo}
\end{equation}
where $W_{ij} = J_{ij} / 4$ and $c_i = (b_i + \sum_j J_{ij})/2$. 
In the following, we will consider QUBO problems expressed as in Eq.~\eqref{eq: qubo}. 
Without loss of generality, we will from now on omit the bias term $\vb{c}$, as it can be accounted for in the quadratic term $W$ at the cost of introducing an additional variable with fixed value $s_{N+1} = 1$.


    \subsection{Quantum annealing}
    The solution of optimization problems using quantum annealing involves mapping the problem onto the ground state of a quantum Hamiltonian, which is then adiabatically prepared~\cite{farhi_quantum_2000, albash_adiabatic_2018}.
    In case of the QUBO problem in Eq.~\eqref{eq: qubo}, the associated quantum Hamiltonian is that of a spin-$\sfrac{1}{2}$ system with Ising interactions
    \begin{equation}
        \hamiltonian_\mathrm{I} = \sum_{i,j = 1}^N W_{ij} \sigmaz{i} \sigmaz{j} \text{,}
        \label{eq: quantum ising hamiltonian}
    \end{equation}
    where $\sigmaz{i}$ is the Pauli-$Z$ operator acting on the $i$-th spin.

    The adiabatic state preparation procedure begins by initializing the system in the ground state of a simple Hamiltonian, typically chosen so that its ground state is known analytically, and then continuously transforms it into $\hamiltonian_\mathrm{I}$ over time.
    According to the adiabatic theorem, if this transformation occurs sufficiently slowly, the system will remain in the instantaneous ground state of the time-dependent Hamiltonian throughout the evolution~\cite{kato_adiabatic_1950}.
    More precisely, we can set $\ket{\psi_0} = \ket{+}^{\otimes N}$ as the initial state at time $t=0$ and implement the time-dependent Hamiltonian
    \begin{equation}
        \hamiltonian(t) = \frac{t}{T} \hamiltonian_\mathrm{I} + \qty(1 - \frac{t}{T}) \hamiltonian_\mathrm{TF} \text{,}
        \label{eq: time-dependent hamiltonian}
    \end{equation}
    where $\hamiltonian_\mathrm{TF} =  - \sum_i \sigmax{i}$ is the transverse field Hamiltonian. 
    At the beginning of the annealing schedule, i.e. at time $t=0$, the state $\ket{\psi_0}$ is the ground state of the initial Hamiltonian $\hamiltonian_\mathrm{TF}$. 
    The system then undergoes adiabatic time evolution, gradually transforming toward the ground state of $\hamiltonian_\mathrm{I}$ as the annealing process reaches the final time $t=T$.
    In the limit of large annealing times $T$, the adiabatic theorem ensures that the system remains in the instantaneous ground state of the Hamiltonian throughout the entire process. 
    The choice of $T$ is therefore crucial for the success of the adiabatic algorithm and should be related to the minimal energy gap $\Delta_{\textrm{min}}$ between the ground state and the first excited state of the Hamiltonian during the evolution.
    Specifically, $T$ should be chosen such that $T \gg 1/\Delta_{\textrm{min}}^2$ to satisfy the adiabatic condition and minimize the probability of diabatic transitions that could lead the system out of the ground state~\cite{farhi_quantum_2000, albash_adiabatic_2018}.
    The final state after the evolution is by construction a product state encoding the solution to the original optimization problem.
    This can be read out by measuring the individual spins in the computational basis.

    While quantum annealing is a promising approach to solving optimization problems, the extent of its advantage over classical algorithms remains under active debate~\cite{albashDemonstrationScalingAdvantage2018, baldassi_efficiency_2018, bauza_scaling_2024, heim_quantum_2015, mandra_deceptive_2018,ronnowDefiningDetectingQuantum2014}.
    Yet, the general agreement is that the potential advantage offered by quantum annealing is intrinsically linked to the formation of entanglement during the evolution according to the time-dependent Hamiltonian~\cite{boixoEvidenceQuantumAnnealing2014, lantingEntanglementQuantumAnnealing2014, albashReexaminationEvidenceEntanglement2015, bauza_scaling_2024}.


    \subsection{Simulated quantum annealing}

    Studies simulating quantum annealing on a classical computer typically rely on techniques such as PIMC or VMC to describe the quantum system~\cite{ronnowDefiningDetectingQuantum2014,santoroTheoryQuantumAnnealing2002,martonakQuantumAnnealingPathintegral2002,hibat-allahVariationalNeuralAnnealing2021a,carleoSimulatingAdiabaticQuantum2024}, as their expressive power allows for the accurate representation of the physical state.
    Here, we approach quantum annealing from a different perspective. 
    We do not seek to simulate the quantum state and its dynamics accurately, but rather to develop a heuristic optimization algorithm that takes inspiration from the quantum annealing process, while being computationally efficient and scalable to large problem sizes. 
    Key to this approach is the choice of an Ansatz that partially trades physical accuracy for computational efficiency, in particular by avoiding the need for Monte Carlo sampling.
    Taking this idea to its logical extreme, recent studies~\cite{bowlesQuadraticUnconstrainedBinary2022,jansen_qudit-inspired_2024,veszeliMeanFieldApproximation2021} proposed approximating the state along the annealing schedule via an Ansatz of product, and thus not entangled, states in the form
    \begin{equation}
        \label{eq: product state ansatz}
        \ket{\psi(\boldsymbol\theta)} = \ket{\theta_1} \otimes \ket{\theta_2} \otimes \ldots \otimes \ket{\theta_N} \text{,}
    \end{equation}
    where each $\ket{\theta_i}$ is a single-spin state fully specified by a real parameter $\theta_i$.
    This method, referred to as \emph{Local Quantum Annealing} (LQA), has the advantage that both the expectation value of the energy along the annealing schedule, and its gradient with respect to the parameters $\boldsymbol{\theta}$, can be computed analytically, without requiring Monte Carlo sampling.
    On the other hand, the Ansatz in Eq.~\eqref{eq: product state ansatz} cannot capture any entanglement between the spins, thus missing the feature believed to be at the core of the quantum annealing's potential advantage.


    \subsection{Generalized Atomic Coherent States}

    The spin-$\sfrac{1}{2}$ product states used by LQA are a subset of the larger class of \emph{Group-Theoretic Coherent States}~\cite{perelomov_coherent_1972, perelomov_generalized_1986}.
    These states allow for the efficient analytical evaluation of expectation values of Pauli operators and their gradient vectors~\cite{guaita_generalization_2021}, but they are limited to the description of unentangled states only.
    
    On the other hand, the ability to efficiently compute expectation values analytically is not exclusive to the product states in Eq.~\eqref{eq: product state ansatz}. 
    Specifically, the class of \emph{Generalized Group-Theoretic Coherent States} has been recently shown to extend that of Group-Theoretic Coherent States by introducing non-trivial correlations between the system's components, while still allowing for a computationally efficient evaluation of the energy and its gradient~\cite{guaita_generalization_2021, schindler_variational_2022,Anders2006,Anders2007}.
    In the following, we will extensively refer to the Generalized Group-Theoretic Coherent States built upon the $\text{SU}\qty(2)$ group for spin-$\sfrac{1}{2}$ particles as \emph{GCS states}.
    A GCS state is defined via the relation $\ket{\psi(x,M,y)} =\nolinebreak \U{y}\V{M}\U{x}\ket{+}^{\otimes N}$, where the operators $\mathcal{U}$ and $\mathcal{V}$ can be expressed as
    \begin{gather}
        \U{x} = \overset{N}{\underset{j=1}{\bigotimes}} \exp(-i \sum_{k\in\qty{x, y, z}} x_{jk} \pauli{k}{j}), \\
        \V{M} = \exp(-i \sum_{j\neq k} M_{jk} \sigmaz{j} \sigmaz{k}) \text{.}
    \end{gather}
    Here, the variational parameters are all real-valued and are grouped into two $N\times3$ matrices, $x$ and $y$, and a symmetric $N\times N$ coupling matrix $M$.
    The total number of parameters is therefore $N_{\mathrm{par}}=6N+N(N-1)/2$.
    Importantly, since $\V{M\equiv 0} = \mathbb{I}$ and any two product states are related by factorized rotations, all product states can be represented exactly by this Ansatz.
    Furthermore, the two-body operators in $\V{M}$ introduce correlations between the spins, generating non-trivial entanglement within the system.

    \begin{figure}[b]
        \centering
        \includegraphics{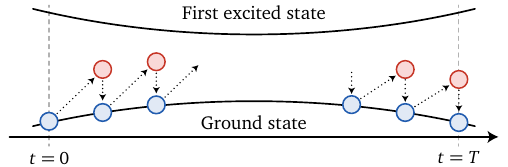}
        \caption{
            \label{fig: optimization scheme}
            \textbf{Sketch of the optimization scheme.}
            As the Hamiltonian evolves, the parameters of the Ansatz are updated to minimize the expectation value of the Hamiltonian. 
            Each time step, only one single update is performed.
        }
    \end{figure}

\begin{figure}[hb]
        \centering
        \includegraphics{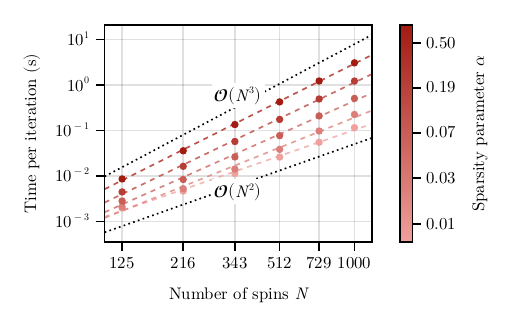}
        \caption{
            \label{fig: execution time}
            \textbf{Time per iteration as a function of the non-zero elements in the adjacency matrix $W$.}
            The number of non-zero elements is varied according to $N_{\mathrm{nnz}} = \qty(1 + \alpha\qty(N-1))N$.
            As the number of non-zero elements is decreased, the execution time per iteration crosses over from $\order{N^3}$ to $\order{N^2}$ scaling.
            The two dotted lines represent guides to the eye for the two scaling regimes. The simulations have been parallelized over $40$ threads.
        }
    \end{figure}

    The optimization scheme we developed initializes the system in a GCS state approximating $\ket{\psi_0} = \ket{+}^{\otimes N}$, with $y\equiv 0$ and $M\equiv 0$.
    Then, the Hamiltonian is varied according to the annealing schedule, evaluating Eq.~\eqref{eq: time-dependent hamiltonian} over a discrete grid of $N_{\mathrm{t}}$ times such that
    \begin{equation}
        \quad t_j < t_{j+1} \, \forall j=1\ldots N_{\mathrm{t}},\quad t_1 = 0 \text{ and } t_{N_{\mathrm{t}}} = T.
    \end{equation}
    The loss function is defined to be the expectation value of the time-dependent Hamiltonian
    \begin{equation}
        \mathcal{L}_{t_j}(x, M, y)=\expval{\hamiltonian(t_j)}{\psi(x,M,y)},
        \label{eq: loss function}
    \end{equation}
    and it is minimized at each time step $t_j$ by the variational algorithm, updating the parameters via gradient-based optimization similarly to Refs. \cite{bowlesQuadraticUnconstrainedBinary2022,veszeliMeanFieldApproximation2021}. 
    We remark that this protocol differs from the unitary dynamics that characterizes the physical quantum annealing process and, from empirical evidence, is more effective in approaching the global minimum when diabatic effects start being relevant. 
    Moreover, under the assumption that the adiabatic evolution results in the state remaining close to the instantaneous ground state along the annealing schedule, we simplify the algorithm by executing only one gradient-based update of the parameters at each time step, as sketched in Fig.~\ref{fig: optimization scheme}. 
    The updates are performed by employing the ADAM optimizer~\cite{kingma_adam_2017}.
    The result is a highly efficient optimization algorithm that allows us to study problems of thousands of variables with ease.
    Specifically, as we show in Sec.~\ref{sec: efficient evaluation of loss}, GCS states allow for the analytical evaluation of expectation values of Pauli operators at a computational cost scaling as $\order{N}$. 
    The number of operations required to evaluate the expectation value of the Hamiltonian, thus, depends on the number of non-zero elements in the adjacency matrix $W$.
    For dense matrices or analogous problems with all-to-all connections, $\order{N^2}$ expectation values need to be computed.
    The total complexity of the algorithm is therefore $\order{N^3}$.
    As the matrix $W$ becomes sparser, on the other hand, the complexity of the algorithm is reduced. 
    In particular, for a sparse matrix $W$ containing $\order{N}$ nonzero elements, the computational cost is reduced to $\order{N^2}$, as shown in Fig.~\ref{fig: execution time}.

    At the end of the annealing schedule, a classical solution to the optimization problem is obtained by evaluating the expectation value $z_j=\expval*{\sigmaz{j}}$ on each spin and rounding it to the closest integer between $-1$ and $1$.
    Note that while the ground state of $\hat H_\mathrm{I}$ is by construction a classical state, the approximate nature of the algorithm and the finite annealing time $T$ inevitably lead to a final state with some residual quantum superposition of classical states. 
    In these cases, the simple protocol to extract a classical solution may lead to discretization errors, and thus represents an additional source of approximation for the optimization algorithm.
    These errors can be mitigated by employing more refined discretization techniques~\cite{montanari_optimization_2021, dupont_extending_2024}, but in the benchmarks presented in this work, we didn't notice a significant improvement from these methods compared to the basic rounding protocol.

    \subsection{Benchmark}

    We test the performance of our algorithm on a set of random instances of the three-dimensional Edwards-Anderson model~\cite{barahonaComputationalComplexityIsing1982}, i.e., a glassy problem defined on a cubic lattice, with couplings drawn from a Gaussian distribution. 
    Due to the cubic topology of the lattice, the adjacency matrix $W$ is sparse and the number of expectation values to be evaluated scales as $\order{N}$, resulting in an overall $\order{N^2}$ scaling of the algorithm's complexity.
    We assess the quality of the solutions by comparing their energies with those from LQA~\cite{bowlesQuadraticUnconstrainedBinary2022}, a standard Simulated Annealing (SA) implementation~\cite{kirkpatrickOptimizationSimulatedAnnealing1983}, and a highly-optimized Parallel Tempering algorithm with Iso-energetic Cluster Moves (PT-ICM)~\cite{bauza_scaling_2024,zhu_efficient_2015}. 
    For each instance, we find the global minimal energies $E_0$ using the Gurobi~\cite{gurobi} exact solver and evaluate the performance based on the relative error $\varepsilon = \qty(E - E_0) / \qty|E_0|$.
    Because of the exponential scaling of Gurobi's runtime, the problem sizes studied in this section are limited to $2197$ spins, corresponding to a cubic lattice with $13$ spins per side.

    \begin{figure}[b]
        \centering
        \includegraphics{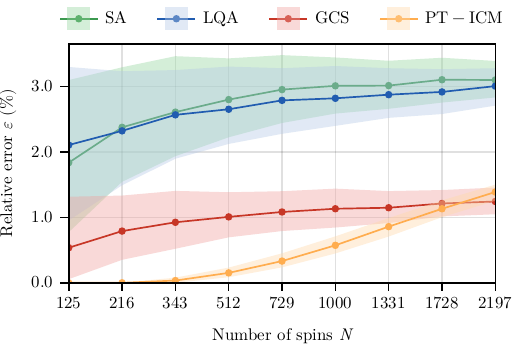}
        \caption{
            \label{fig: relative error vs system size}
            \textbf{Median relative error $\varepsilon$ as a function of the number of spins $N$.}
            Each method is run for $N_{\mathrm{t}} = 1000$ iterations on $1000$ random problem instances.
            The shaded regions represent the interquartile ranges of the resulting distributions.
        }
    \end{figure}

    Fig.~\ref{fig: relative error vs system size} shows the evolution of the relative error $\varepsilon$ as a function of the number of spins $N$. 
    Each algorithm is executed on $1000$ random problem instances, and the median of $\varepsilon$ over these runs is reported after $N_{\mathrm{t}} = 1000$ iterations. For GCS and LQA, a single iteration corresponds to one update of the variational parameters, whereas for SA and PT-ICM, an iteration is defined as one full Monte Carlo sweep over all spins
    The shaded regions represent the interquartile range of the resulting distributions, defined as the range between the $25\%$ and the $75\%$ of the distribution, thus covering the middle $50\%$ of the data.
    We observe that, while SA and LQA achieve comparable error rates, our algorithm consistently outperforms them by a significant margin.
    When comparing it to PT-ICM, on the other hand, we find that for small system sizes the latter returns lower-energy solutions on average. 
    However, as the number of spins increases, the performance of PT-ICM quickly deteriorates, while GCS displays a more gradual increase in the relative error.
    For the largest system sizes considered, GCS outperforms all other methods tested.

    \begin{figure*}[t]
        \centering
        \includegraphics{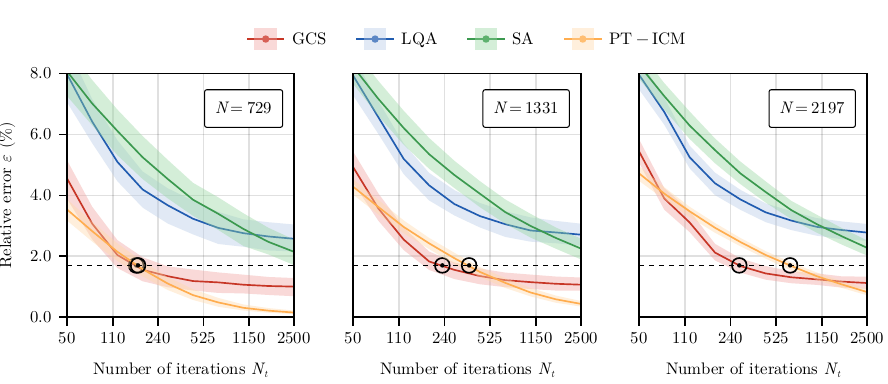}
        \caption{
            \label{fig: relative error vs iteration number}
            \textbf{Median relative error $\varepsilon$ as a function of the number of iterations $N_{\mathrm{t}}$.} 
            Each method is benchmarked on $1000$ random problems.
            The shaded regions represent the interquartile range of the resulting distributions.
            We select a target error threshold (identified by the dashed line) and analyze the number of iterations required by GCS and PT-ICM to reach it (marked by the black dots).
        }
    \end{figure*}

    In Tab.~\ref{tab: runtimes} we present the execution times for the four heuristics under comparison, alongside those of the Gurobi solver.
    We emphasize that both PT-ICM and Gurobi are highly optimized algorithms, whereas our self-implemented versions of SA, LQA and GCS, despite being carefully developed, may benefit from further optimization.
    Therefore, we argue that execution times provide only partial insight into the relative performance of the heuristics.
    A more informative comparison correlates the algorithms' results to the number of iterations $N_{\mathrm{t}}$ they have been executed for, while accounting for the distinct complexity scaling of each heuristic. 

    \begin{table}[h]
        \centering
        \begin{tabular}{c||c|c|c|c|c}
            $N$ & \multicolumn{5}{|c}{Runtime (s)} \\
                & SA & LQA & GCS & PT-ICM & Gurobi \\
            \hline
            $125$ & $ 0.01 $ & $ 0.01 $ & $ 2.21 $ & $ 0.18 $ & $ 0.10 \; \qty(0.09, \, 0.13) $ \\
            $216$ & $ 0.02 $ & $ 0.02 $ & $ 4.44 $ & $ 0.32 $ & $ 0.62 \; \qty(0.46, \, 0.93) $ \\
            $343$ & $ 0.03 $ & $ 0.03 $ & $ 9.71 $ & $ 0.50 $ & $ 2.91 \; \qty(2.61, \, 3.22) $ \\
            $512$ & $ 0.05 $ & $ 0.04 $ & $ 20.28 $ & $ 0.75 $ & $ 8.67 \; \qty(7.25, \, 10.06) $ \\
            $729$ & $ 0.07 $ & $ 0.05 $ & $ 38.17 $ & $ 1.06 $ & $ 25.55 \; \qty(22.54, \, 29.26) $ \\
            $1000$ & $ 0.10 $ & $ 0.07 $ & $ 69.71 $ & $ 1.46 $ & $ 70.53 \; \qty(58.08, \, 89.57) $ \\
            $1331$ & $ 0.13 $ & $ 0.09 $ & $ 124.59 $ & $ 1.95 $ & $ 157.50 \; \qty(127.34, \, 251.59) $ \\
            $1728$ & $ 0.17 $ & $ 0.12 $ & $ 207.72 $ & $ 2.53 $ & $ 664.59 \; \qty(423.89, \, 1353.82) $ \\
            $2197$ & $ 0.22 $ & $ 0.16 $ & $ 449.56 $ & $ 3.20 $ & $ 1568.23 \; \qty(934.55, \, 3125.83) $ \\
        \end{tabular}
        \caption{
            \label{tab: runtimes}
            \textbf{Median runtimes of the tested algorithms as a function of the number of spins $N$.}
            Runtimes of SA, LQA, GCS and PT-ICM instances showed negligible variance. 
            Gurobi's runtime instead highly depends on the problem instance, resulting in a broad and skewed distribution.
            In this case, we reported within brackets the $25\textsuperscript{th}$ and $75\textsuperscript{th}$ percentiles in addition to the median.
            All simulations of GCS have been parallelized over $40$ threads, while Gurobi over $18$ of them.
        }
    \end{table}

\begin{figure}[hb]
        \centering
        \includegraphics{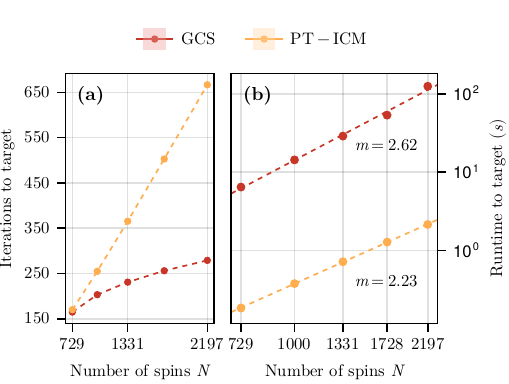}
        {\phantomsubcaption \label{fig: iterations to target}}
        {\phantomsubcaption \label{fig: runtime to target}}
        \caption{
            \label{fig: runtime scaling}
            \textbf{Resource analysis to reach a given error threshold.}
            \textbf{a)}~--~Number of iterations required by GCS and PT-ICM to reach an error threshold of $1.7\%$ as a function of the system size. 
            \textbf{b)}~--~Total runtime required by GCS and PT-ICM to reach an error threshold of $1.7\%$ as a function of the system size. 
            Dashed lines indicate power-law fits to the data, with the corresponding scaling exponents annotated next to each line.
            GCS simulations have been parallelized over $40$ threads on a single computer.}
    \end{figure}

    As reported in Tab.~\ref{tab: runtimes}, it is important to acknowledge that one single iteration of GCS is computationally more expensive than one iteration of PT-ICM. 
    As all heuristics perform better when the number of iterations is increased, a fair comparison should account for this difference by adjusting $N_{\mathrm{t}}$ for each method accordingly.
    In Fig.~\ref{fig: relative error vs iteration number}, we present the relative error $\varepsilon$ as a function of the number of iterations $N_{\mathrm{t}}$, for different system sizes, providing a complementary evaluation of the algorithms' performance.
    As expected for simulated classical annealing methods, the relative error of SA and PT-ICM decreases steadily with $N_{\mathrm{t}}$.
    In contrast, GCS asymptotically approaches a constant value. 
    We argue that this constant value is due to the limited expressivity of the Ansatz.
    The similarity of LQA's trend to GCS supports this hypothesis, with LQA converging to a higher error level due to the absence of entanglement in the product state Ansatz.
    The rapid convergence of GCS allows us to identify a tolerated error level near the elbow of the curve, and to study the runtime required by the algorithms to reach it.
    For this analysis we focus on the two best-performing algorithms, i.e., GCS and PT-ICM. 
    The dashed line in Fig.~\ref{fig: relative error vs iteration number} denotes the chosen error threshold ($1.7\%$), while black dots mark its intersections with the GCS and PT-ICM median results.
    Notably, while both algorithms require an increasing number of iterations to reach the target accuracy as the system size grows, PT-ICM exhibits an approximately linear scaling with system size, whereas GCS displays a sublinear trend. 
    This behavior is illustrated in Fig.~\ref{fig: iterations to target}, where the number of iterations required to reach the error threshold is plotted as a function of system size.

    Building on this, we analyze the total runtime needed by each algorithm to reach the target error, accounting for both the number of required iterations and their individual cost.
    The results are presented in Fig.~\ref{fig: runtime to target}, which shows the total runtime as a function of the number of spins. Power-law fits to the data reveal that, despite the per-iteration complexity being quadratic for GCS and linear for PT-ICM, the total runtime scaling to reach the threshold is comparable for both algorithms, falling between quadratic and cubic.
    Specifically, we find that the total runtime scales as $\mathcal{O}(N^\alpha)$, with $\alpha_{\mathrm{GCS}} = 2.62$ and $\alpha_{\mathrm{PT-ICM}} = 2.23$.
    The error on the power $\alpha$ has been estimated via bootstrapping and verified to be negligible.

    While the runtime scaling provides insight into the algorithmic efficiency in the large system size limit, practical performance is also determined by the absolute prefactors -- i.e., the intercepts in \cref{fig: runtime to target}. 
    In our benchmarks, we find that the average runtime of PT-ICM is approximately $40$ times shorter than that of GCS for the same system size.
    However, unlike the scaling exponents, these prefactors strongly depend on implementation-specific aspects rather than the intrinsic properties of the evaluated algorithms. 
    Factors such as the choice of programming language, the extent of code optimization, and the potential for parallel execution all contribute significantly to the observed offsets.
    The GCS algorithm was implemented in Julia with basic performance considerations, though it remains open to further tuning.
    In contrast, the PT-ICM implementation is written in Rust -- a compiled language renowned for its high performance -- and is heavily optimized.
    Regarding parallelization, GCS naturally allows concurrent execution across multiple threads and computing nodes via distributed evaluation of the gradients.
    In the simulations presented here, we exploited parallelism over $40$ threads within a single node, though the algorithm’s design also supports distributed computing.
    Conversely, PT-ICM relies on a parallel tempering framework in which Markov chains at different temperatures are evolved simultaneously. 
    While this approach allows for some concurrency, practical optimization implementations typically use only a limited number of chain replicas, constraining the overall effectiveness of parallelization.
    As a result, the reported runtime prefactors should be interpreted with caution, as they reflect not only the intrinsic efficiency of the algorithms but also differences in implementation maturity and execution environments.

    The results presented in \cref{fig: relative error vs system size,fig: relative error vs iteration number,fig: runtime scaling} collectively demonstrate that, for fixed annealing times and in the limit of large system sizes, the GCS algorithm outperforms all other methods tested in this study. 
    When the number of iterations is not fixed, GCS converges to its asymptotic error level faster than the typical convergence time of the other algorithms, requiring fewer iterations to reach a given error threshold. 
    The consequence is that, despite their different iteration costs, the overall runtime scaling of GCS and PT-ICM to reach such threshold is similar, making GCS a competitive alternative to classical annealing methods.


\section{Discussion}

    In this work, we introduced a novel quantum-inspired heuristic for solving QUBO problems.
    We developed a variational procedure inspired by the quantum annealing process, for which the loss function and its gradient can be computed analytically, allowing for the optimization of problems with thousands of variables within minutes.
    The variational Ansatz, based on Generalized Group-Theoretic Coherent States, partially captures the entanglement emerging during the quantum annealing dynamics, leading to an approximate solution outperforming a corresponding uncorrelated Ansatz.
    Concurrently, the analytical evaluation of the expectation values of the Hamiltonian and its gradient allows for a highly scalable algorithm with a low-degree polynomial complexity in the number of spins.

    We demonstrate the effectiveness of our algorithm on the 3D Edward-Anderson model, through comprehensive benchmarking against established optimization methods. 
    Our results clearly identify a wide range of cases where GCS contends with all other methods tested in terms of the quality of the solutions found, including the Parallel Tempering algorithm with Iso-energetic Cluster Moves.
    In particular, we show that for a fixed number of iterations and in the limit of large system size, the GCS algorithm shows the best performance. 
    Moreover, by fixing a tolerated error threshold, we analyze the runtime scaling of our algorithm and demonstrate that it grows similarly to that of the state-of-the-art PTICM.
    
    Several avenues for further research remain open. 
    Increasing the Ansatz's expressivity, for example, could potentially lead to a more accurate representation of the quantum annealing dynamics, thus improving the quality of the solutions.
    This could be achieved by employing linear superpositions of GCS states. Indeed, if the states in the superposition share the same values of the $y$ and $M$ parameters, differing only in the $x$ parameters, the Ansatz preserves the possibility to compute the loss function and its gradient analytically. 

    In the opposite direction, one could consider simplifying the Ansatz by enforcing structures on the parameters, thus reducing their number and the complexity of the optimization problem.
    A prototypical case would be the invariance of the problem under specific symmetries, which could be readily incorporated into the Ansatz, eventually accelerating its optimization procedure.
    In the case of problems with a known topology, like the one studied here, one could also consider a sparse matrix $M$ bearing the same structure as the adjacency matrix $W$. 
    This would reduce the expressivity of the Ansatz, but still incorporate entanglement while considerably speeding up the optimization algorithm compared to the GCS with an arbitrary parameter matrix $M$.
    
    In conclusion, the results presented in this work demonstrate the potential of quantum-inspired heuristics for solving large-scale optimization problems. 
    Our comparison against simpler Ans\"atze shows that the entanglement captured by the GCS framework plays a crucial role in improving the quality of the solutions.
    Our work thus paves the way for leveraging quantum effects for complex classical optimization problems, potentially opening new avenues for the development of efficient algorithms for large-scale optimization tasks.


\section{Methods\label{sec: efficient evaluation of loss}}

At the core of our algorithm's efficiency lies the analytical evaluation of both the expectation value of the Hamiltonian and its gradient. 
In this section, we provide a concise overview of the method employed to evaluate the loss function.
For a more comprehensive explanation, as well as a similar treatment of the gradient, we direct the reader to the Supplementary Note 1.

We begin by noting that the loss function in Eq.~\eqref{eq: loss function} is expressed as a  linear combination of $1$- and $2$-local terms in the form of $\expval*{\sigmax{i}}$ and $\expval*{\sigmaz{i}\sigmaz{j}}$, respectively.
Here, we detail the procedure used to estimate $\expval*{\sigmax{i}}$ for a specific spin index $i$.
The evaluation of $\expval*{\sigmaz{i}\sigmaz{j}}$ follows an analogous approach, leading to similar results.

Let us introduce the notation $\pauli{\alpha}{i}$ (using a Greek index) to represent the Pauli operators $\sigma_{-}$, $\sigma_{z}$ and $\sigma_{+}$ for $\alpha\in\qty{-1, 0, 1}$, respectively.
Additionally, we define the states
\begin{gather}
    \ket{\psi(x)} = \U{x}\ket{+}^{\otimes N} \text{,}\\
    \ket{\psi(x,M)} = \V{M}\U{x}\ket{+}^{\otimes N} 
\end{gather}
to simplify our notation.
The operator $\V{M}$ obeys the following relation, reported in~\textcite{guaita_generalization_2021}:
\begin{equation}
    \V{M}^\dagger\pauli{\alpha}{i} \V{M} =  \exp\Big(-4i\alpha \sum_{j=1}^N M_{ij} \sigmaz{j}\Big)\pauli{\alpha}{i} \label{eq: commutation V} \text{.}
\end{equation}
Similarly, the operator $\U{y}$ satisfies 
\begin{equation}
    \U{y}^\dagger \pauli{j}{i} \U{y} = \sum_{\alpha\in\qty{-1, 0, 1}} d_{j\alpha}^{(i)}(y) \pauli{\alpha}{i} \text{,}
        \label{eq: commutation U}
\end{equation}
where the coefficients $d_{j\alpha}^{(i)}$ are analytical functions of the parameters $y$.

By definition, the expectation value $\expval*{\sigmax{i}}$ is given by
\begin{equation}
    \expval*{\sigmax{i}} = \expval*{\U{y}^\dagger \sigmax{i}\U{y}}{\psi(x, M)}\text{.}
\end{equation}
Applying Eq.~\eqref{eq: commutation U} to the above expression, we obtain
\begin{align}
    \expval*{\sigmax{i}} &= \sum_{\alpha\in\qty{-1, 0, 1}} d_{x\alpha}^{(i)}(y) \expval*{\pauli{\alpha}{i}}{\psi(x, M)}   \nonumber \\
    &= \sum_{\alpha\in\qty{-1, 0, 1}} d_{x\alpha}^{(i)}(y) \expval*{\V{M}^\dagger\pauli{\alpha}{i}\V{M}}{\psi(x)} \text{.}
\end{align}
Next, we use Eq.~\eqref{eq: commutation V} to rewrite the expectation value as
\begin{align}
    \expval*{\sigmax{i}} &= \sum_{\alpha\in\qty{-1, 0, 1}} d_{x\alpha}^{(i)}(y) \times \nonumber \\
    &\times \expval*{\exp\Big(-i4\alpha \sum_{j=1}^N M_{ij} \sigmaz{j}\Big)\pauli{\alpha}{i} }{\psi(x)} \text{.}
\end{align}
Finally, we decompose the state $\ket{\psi(x)} = \otimes_{j=1}^N \ket{\psi(x_j)}$ and use it to express the expectation value as
\begin{align}
    \expval*{\sigmax{i}} &= \sum_{\alpha\in\qty{-1, 0, 1}} d_{x\alpha}^{(i)}(y) \times \nonumber \\
    &\times \prod_{j=1}^N \expval*{ \exp\Big(-4i\alpha M_{ij} \sigmaz{j}\Big) (\pauli{\alpha}{i})^{\delta_{ij}}}{\psi(x_j)} \text{.}
    \label{eq: factorized expval}
\end{align}
Each of the $N$ terms in the product can now be computed analytically with a number of operations that remains independent of the total number of spins. 
Thus, the overall computational complexity of evaluating the expectation value in Eq.~\eqref{eq: factorized expval} scales as $\order{N}$.

An analogous procedure leads to similar results for the evaluation of $\expval*{\sigmaz{i}\sigmaz{j}}$ and the gradient vectors (see Supplementary Note 1).


\section*{acknowledgments}
We acknowledge several fruitful discussions with Alberto Mercurio, Filippo Ferrari, Luca Gravina, and Alessandro Sinibaldi.
This work was supported by the Swiss National Science Foundation through Project No. 200020\_215172, and was supported as a part of NCCR SPIN, a National Centre of Competence in Research, funded by the Swiss National Science Foundation (grant number 225153).


\section*{Data availability}

The data that support the findings of this study are available from the corresponding author upon reasonable request.

\section*{Code availability}

The code used to generate the results presented in this work is openly available on GitHub at \url{https://github.com/LorenzoFioroni/QuboSolver.jl}.


\section*{Author contributions}
V.S. conceived the project. L.F. developed the code and performed the numerical simulations. Both authors contributed equally to the analysis and interpretation of the results and to the writing of the manuscript.


\section*{Competing interests}
The authors declare no competing interests.


\bibliography{bibliography}

\clearpage
\onecolumngrid


\section*{Supplementary Note 1}


\subsection{\label{supmat sec: efficient computation of the loss function} Efficient computation of the loss function}

The framework we developed allows for the efficient analytical evaluation of both the expectation value of the time-dependent Hamiltonian 
\begin{equation}
    \hamiltonian(t) = \frac{t}{T} \hamiltonian_\mathrm{I} + \qty(1 - \frac{t}{T}) \hamiltonian_\mathrm{TF} \text{,}
\end{equation} 
and its gradient.
This is achieved by leveraging the commutation properties of the operators $\U{x}$ and $\V{M}$, as described in~\textcite{guaita_generalization_2021}.

First, note that the Hamiltonian consists of a sum of $1$- and $2$-local terms in the form of $\expval*{\sigmax{i}}$ and $\expval*{\sigmaz{i}\sigmaz{j}}$, respectively.
Each of these terms can be evaluated independently, allowing for the full evaluation of the loss to be carried out in parallel.
In this section, we detail the procedure for estimating $\expval*{\sigmax{i}}$ for a given spin index $i$.
The evaluation of $\expval*{\sigmaz{i}\sigmaz{j}}$ follows a similar approach, which we will briefly outline at the end.
    
The goal is to express $\expval*{\sigmax{i}}$ as a linear combination of expectation values of separable operators on product states, which can be evaluated analytically with linear complexity in the number of spins.
Let $\pauli{\alpha}{i}$ (with a Greek index) denote the Pauli operators $\sigma_{-}$, $\sigma_{z}$ and $\sigma_{+}$ for $\alpha\in\qty{-1, 0, 1}$, respectively.
The operators $\V{M}$ and $\U{y}$ obey the following relations:
\begingroup
    \addtolength{\jot}{1em}
    \begin{gather}
        \V{M}^\dagger\pauli{\alpha}{i} \V{M} = \exp\Big(-4i\alpha \sum_{j=1}^N M_{ij} \sigmaz{j}\Big)\pauli{\alpha}{i}  \label{supmat eq: commutation V} \text{,} \\
        \U{y}^\dagger\pauli{j}{i} \U{y} = \sum_{k\in\qty{x, y, z}} c_{jk}^{(i)}(y) \pauli{k}{i} \text{.} \label{supmat eq: commutation U}
    \end{gather}
\endgroup
The coefficients $c_{jk}^{(i)}$ in Eq.~\eqref{supmat eq: commutation U} are analytical functions depending solely on the vector of parameters $\vb{y}_{i}$, and can be evaluated efficiently~\cite{guaita_generalization_2021}.
Next, we introduce the matrix $A$ that transforms the Pauli basis $\{\pauli{j}{i}\}_{j,i}$ to the basis $\{\pauli{\alpha}{i}\}_{\alpha,i}$.
Consequently, Eq.~\eqref{supmat eq: commutation U} can be rewritten as
\begin{equation}
    \U{y}^\dagger \pauli{j}{i} \U{y} = \sum_{\alpha\in\qty{-1, 0, 1}} d_{j\alpha}^{(i)}(y) \pauli{\alpha}{i} \text{,}
    \label{supmat eq: commutation U complexified}
\end{equation}
where $d_{j\alpha}^{(i)}(y) = \sum_{k\in\qty{x, y, z}} c_{jk}^{(i)}(y) A_{k \alpha}$.
Finally, we define the following states for brevity of notation
\begin{equation}
    \ket{\psi(x)} = \U{x}\ket{+}^{\otimes N} \quad \text{and} \quad \ket{\psi(x, M)} = \V{M}\U{x}\ket{+}^{\otimes N} \text{.}
\end{equation}

Using Eq.~\eqref{supmat eq: commutation U complexified}, the expectation value of $\sigmax{i}$ can be expressed as 
\begingroup
    \addtolength{\jot}{1em}
    \begin{align}
            \expval*{\sigmax{i}} & = \expval*{\U{y}^\dagger \sigmax{i} \U{y}}{\psi(x, M)}  \\
        & = \sum_{\alpha\in\qty{-1, 0, 1}} d_{x\alpha}^{(i)}(y) \expval*{ \V{M}^\dagger \pauli{\alpha}{i} \V{M} }{\psi(x)} \text{.}
    \end{align}
\endgroup
Next we apply the relation in Eq.~\eqref{supmat eq: commutation V}, which yields
\begin{equation} 
    \expval*{\sigmax{i}}  = \sum_{\alpha\in\qty{-1, 0, 1}} d_{x\alpha}^{(i)}(y)  \expval*{\exp\Big(-4i\alpha \sum_{j=1}^N M_{ij} \sigmaz{j}\Big) \pauli{\alpha}{i} }{\psi(x)} \text{.}
\end{equation}
Finally, we observe that $\ket{\psi(x)}$ is the product state
\begin{equation}
    \ket{\psi(x)} = \U{x}\ket{+}^{\otimes N} = \underset{j=1}{\overset{N}{\otimes}} \exp(-i \vb{x}_j \cdot \boldsymbol{\sigma}^{(j)}) \ket{+}  = \underset{j=1}{\overset{N}{\otimes}} \ket{\psi(x_j)}\text{.} \label{supmat eq: factorized psi_x}
\end{equation}
Employing the relation in Eq.~\eqref{supmat eq: factorized psi_x}, the expectation value is rewritten as a factorized product of single-spin expectation values, and can therefore be evaluated efficiently in time $\order{N}$:
\begin{equation} 
    \label{supmat eq: expval sigma_x}
    \expval{\sigmax{i}} = \sum_{\alpha\in\qty{-1, 0, 1}} d_{x\alpha}^{(i)}(y) \prod_{j=1}^N\expval*{ \exp\Big(-4i\alpha M_{ij} \sigmaz{j}\Big) (\pauli{\alpha}{i})^{\delta_{ij}} }{\psi(x_j)} \text{.}
\end{equation}

A similar procedure can be followed to evaluate the expectation value of $2$-local operators $\expval*{\sigmaz{i}\sigmaz{j}}$.
We now use, in addition to Eq.~\eqref{supmat eq: commutation U}, the unitary relation $\U{x}\U{x}^\dagger = \mathbb{I}$ to get
\begingroup
    \addtolength{\jot}{1em}
    \begin{align}
        \expval*{\sigmaz{i}\sigmaz{j}} & = \expval*{\U{y}^\dagger \sigmaz{i}  \sigmaz{j} \U{y}}{\psi(x, M)} \\
        & = \expval*{\U{y}^\dagger \sigmaz{i} \, \U{y} \U{y}^\dagger \, \sigmaz{j} \U{y}}{\psi(x, M)} \\
        & = \sum_{\alpha,\beta\in\qty{-1, 0, 1}}  d_{z\alpha}^{(i)}(y) d_{z\beta}^{(j)}(y) \expval*{ \V{M}^\dagger \pauli{\alpha}{i} \pauli{\beta}{j} \V{M} }{\psi(x)} \text{.}
    \end{align}
\endgroup
Similarly, we employ Eq.~\eqref{supmat eq: commutation V} as well as $\V{M}\V{M}^\dagger = \mathbb{I}$ to obtain
\begingroup
    \addtolength{\jot}{1em}
    \begin{align}
        \expval*{\sigmaz{i}\sigmaz{j}} &= \sum_{\alpha,\beta\in\qty{-1, 0, 1}}  d_{z\alpha}^{(i)}(y) d_{z\beta}^{(j)}(y) \expval*{  \exp(-4i\alpha \sum_{k=1}^N  M_{ik} \sigmaz{k}) \pauli{\alpha}{i} \exp(-4i\beta \sum_{k=1}^N M_{jk} \sigmaz{k}) \pauli{\beta}{j} }{\psi(x)} \\
        &= \sum_{\alpha,\beta\in\qty{-1, 0, 1}} \!\!\! d_{z\alpha}^{(i)}(y) d_{z\beta}^{(j)}(y) \prod_{k=1}^N \expval*{ \exp(-4i\alpha M_{ik} \sigmaz{k}) (\pauli{\alpha}{i})^{\delta_{ik}} \exp(-4i\beta  M_{jk} \sigmaz{k})(\pauli{\beta}{j})^{\delta_{jk}} }{\psi(x_k)} \text{,}
    \end{align}
\endgroup
where we used the factorized expression for $\ket{\psi(x)}$ in the second step.
Note that, although the sums now run over nine possible combinations of $\alpha$ and $\beta$, the scaling of the computation with respect to the number of spins remains linear.


\subsection{\label{supmat sec: efficient computation of the gradient vector} Efficient computation of the gradient vector}

A procedure akin to the one outlined in the previous section allows for the efficient computation of the gradient of the loss function with respect to the variational parameters.
Analogously to our previous discussion, we focus on the expectation value $\expval*{\sigmax{i}}$ of a local operator and show that its gradient vector can be evaluated analytically with an $\order{N}$ complexity.
A similar result can be equivalently proven for $\expval*{\sigmaz{i}\sigmaz{j}}$.

Let us denote with $P_{ij}^\alpha$ the $j$-th element of the product in Eq.~\eqref{supmat eq: expval sigma_x}
\begin{equation}
    P_{ij}^\alpha(x, M) = \expval*{  \exp\Big(-4i\alpha M_{ij} \sigmaz{j}\Big) (\pauli{\alpha}{i})^{\delta_{ij}}}{\psi(x_j)} \text{.}
\end{equation}
The expectation value in Eq.~\eqref{supmat eq: expval sigma_x} can thus be  written as
\begin{equation} 
    \label{supmat eq: expval sigma_x compact}
    \expval*{\sigmax{i}}  = \sum_{\alpha\in\qty{-1, 0, 1}} d_{x\alpha}^{(i)}(y) P_i^\alpha (x, M) \text{,} 
\end{equation}
where we defined $P_i^\alpha(x, M) = \prod_j P_{ij}^\alpha(x, M)$.
Note that the evaluation of $P_{ij}^\alpha(x, M)$ is efficient and can be carried out with $\order{1}$ complexity.

We first discuss the computation of the gradient with respect to the $x$ parameters.
The only dependence of $\expval*{\sigmax{i}}$ on the parameter $x_{lm}$ is through the single-spin state $\ket{\psi(x_l)}$ which, in turn, only appears in $P_{il}^\alpha$.
Consequently, the derivative of $\expval*{\sigmax{i}}$ with respect to $x_{lm}$ can be written as
\begin{equation}
    \pdv{\expval*{\sigmax{i}}}{x_{lm}} = \sum_{\alpha\in\qty{-1, 0, 1}} d_{x\alpha}^{(i)}(y) \pdv{ P_{il}^\alpha}{x_{lm}} \prod_{j\neq l} P_{ij}^\alpha = \sum_{\alpha\in\qty{-1, 0, 1}} d_{x\alpha}^{(i)}(y) \frac{P_{i}^\alpha}{P_{il}^\alpha} \pdv{ P_{il}^\alpha}{x_{lm}} 
    \label{supmat eq: implicit grad_x}
\end{equation}
Let us now define the derived state $\ket{\partial_{x_{lm}}\psi(x_l)} = \pdv{\ket{\psi(x_l)}}{x_{lm}}$.
We can finally express the derivative $\partial P_{il}^\alpha / x_{lm}$ as
\begin{equation}
    \pdv{ P_{il}^\alpha}{x_{lm}} = \bra{\partial_{x_{lm}} \psi(x_l)}  \exp\Big(-4i\alpha M_{il} \sigmaz{l}\Big)(\pauli{\alpha}{i})^{\delta_{il}}\ket{\psi(x_l)} + \bra{\psi(x_l)} \exp\Big(-4i\alpha M_{il} \sigmaz{l}\Big)(\pauli{\alpha}{i})^{\delta_{il}} \ket{ \partial_{x_{lm}}\psi(x_l)} \text{.} \label{supmat eq: derivative P_x}
\end{equation}
Notice that Eq.~\eqref{supmat eq: derivative P_x}, and therefore the derivative $\partial \expval*{\sigmax{i}} / \partial x_{lm}$ can be evaluated with $\order{1}$ complexity.
Thus, the computation of the full gradient vector with respect to the $x$ parameters $\nabla_x \expval*{\sigmax{i}}$ has a linear cost in the number of spins.

Following a similar procedure, we observe that $\expval*{\sigmax{i}}$ depends on the parameter $M_{lm}$ only if $l=i$. 
We thus restrict to this case and obtain 
\begin{equation}
    \pdv{\expval*{\sigmax{i}}}{M_{im}} = \sum_{\alpha\in\qty{-1, 0, 1}} d_{x\alpha}^{(i)}(y) \pdv{ P_{im}^\alpha}{M_{im}} \prod_{j\neq i} P_{ij}^\alpha = \sum_{\alpha\in\qty{-1, 0, 1}} d_{x\alpha}^{(i)}(y) \frac{P_{i}^\alpha}{P_{ii}^\alpha} \pdv{ P_{im}^\alpha}{M_{im}} \text{,}
    \label{supmat eq: implicit grad_M}
\end{equation}
where $\partial { P_{im}^\alpha} / \partial {M_{im}}$ is found to be
  \begin{equation}
    \pdv{ P_{im}^\alpha}{M_{im}} = -4i\alpha  \bra{\psi(x_m)}  \exp\Big(-4i\alpha M_{im} \sigmaz{m}\Big) \sigmaz{m} \; (\pauli{\alpha}{i})^{\delta_{im}}  \ket{\psi(x_m)} \text{.} \label{supmat eq: derivative P_M}
\end{equation}
Also in this case the computation of each derivative has constant cost in the number of spins.
Since the only parameters that can yield a non-zero derivative are $\qty{M_{im}}_{m=1}^N$, the computation of the full gradient vector $\nabla_M \expval*{\sigmax{i}}$ has $\order{N}$ complexity.

Finally, we turn our focus to the $y$ parameters.
As previously mentioned, the coefficients $d_{x\alpha}^{(i)}(y)$ only depend on the $\vb{y}_i$ vector of parameters.
Consequently, we anticipate $\partial{\expval*{\sigmax{i}}} / \partial {y_{lm}} = 0$ for $l\neq i$.
We can therefore compute the derivative of $\expval*{\sigmax{i}}$ with respect to the parameter $y_{im}$, finding
\begin{equation}
    \pdv{\expval*{\sigmax{i}}}{y_{im}}  = \sum_{\alpha\in\qty{-1, 0, 1}} \pdv{d_{x\alpha}^{(i)}(y)}{y_{im}} \prod_{j=1}^N P_{ij}^\alpha(x, M) \text{.} 
    \label{supmat eq: grad_y}
\end{equation}
The derivatives $\partial{d_{x\alpha}^{(i)}(y)} / \partial {y_{im}} $ are analytical functions of the $y$ parameters, similar to $d_{x\alpha}^{(i)}(y)$.
The expression in Eq.~\eqref{supmat eq: grad_y} can be evaluated with $\order{N}$ complexity.
Since $\qty{y_{im}}_{m=1}^3$ are the only parameters that can yield a non-zero derivative, the complexity for the computation of the full gradient vector remains unchanged and scales linearly with the number of spins.

\end{document}